# Perpendicularly Magnetized Ni / Pt (001) Epitaxial Superlattice

T. Seki,[1,2,3,*] M. Tsujikawa,[2,4] K. Ito,[1,2] K. Uchida,[1,2,3] H. Kurebayashi,[5]

M. Shirai,[2,4,6] and K. Takanashi[1,2,6]

[1] *Institute for Materials Research, Tohoku University, Sendai 980-8577, Japan*

[2] *Center for Spintronics Research Network, Tohoku University, Sendai 980-8577, Japan*

[3] *National Institute for Materials Science, Tsukuba 305-0047, Japan*

[4] *Research Institute of Electrical Communication, Tohoku University, Sendai 980-8577, Japan*

[5] *London Centre for Nanotechnology and Department of Electronic & Electrical Engineering, UCL, WC1H 0AH, UK*

[6] *Center for Science and Innovation in Spintronics, Core Research Cluster, Tohoku University, Sendai 980-8577, Japan*

\* e-mail: go-sai@imr.tohoku.ac.jp

**Abstract**

A perpendicularly magnetized ferromagnetic layer is an important building block for recent/future high-density spintronic memory applications. This paper reports on the fabrication of perpendicularly magnetized Ni / Pt superlattices and the characterization of their structures and magnetic properties. The optimization of film growth conditions allowed us to grow epitaxial Ni / Pt (001) superlattices on $SrTiO_3$ (001) single crystal substrates. We investigated their structural parameters and magnetic properties as a function of the Ni layer thickness, and obtained a high uniaxial magnetic anisotropy energy of $1.9 \times 10^6$ erg/cm$^3$ for a [Ni (4.0 nm) / Pt (1.0 nm)] superlattice. In order to elucidate the detailed mechanism on perpendicular magnetic anisotropy for the Ni / Pt (001) superlattices, the experimental results were compared with the first-principles calculations. It has been found that the strain effect is a prime source of the emergence of perpendicular magnetic anisotropy.

# 1. Introduction

A ferromagnetic layer exhibiting room-temperature perpendicular magnetization is an important building block for the development of various spintronic applications [Refs.1-9] such as ultrahigh density magnetic recording devices, magnetic random access memories, and three-terminal spintronic devices. The perpendicularly magnetized state at zero external magnetic field is achieved when a ferromagnetic layer possesses a magnetic anisotropy field in the normal direction to the film plane larger than the demagnetizing field. This perpendicular magnetic anisotropy (PMA) can be obtained by utilizing the bulk magnetic anisotropy and the interface magnetic anisotropy. The metallic superlattices such as Co / Pd [Refs.10-12], Co / Pt [Refs.11-13], and Co / Ni [Refs.14-19] are the materials systems showing clear PMA due to the interface magnetic anisotropy, and several origins for the emergence of interface magnetic anisotropy have been proposed [Ref.20]: (i) interface magnetocrystalline anisotropy due to the anisotropic atomic configuration at the interface, which is known as the Néel model, (ii) interface magnetoelastic anisotropy due to the anisotropic strain originating from the lattice mismatch, and (iii) alloying at the interface.

Ni-based metallic superlattices are a representative of PMA originating from a lattice strain, and a recent study has shown electric-field-driven switching of magnetization for the Cu / Ni multilayers through the magnetoelectric coupling effect [Ref.21]. Pioneering works on the magnetic anisotropy for the ultrathin Ni [Refs.22,23] reported that the Ni layer grown on Cu (001) was spontaneously magnetized in the perpendicular direction to the film plane at a certain layer thickness. The perpendicular magnetization of the Ni layer was attributable to the strain-induced perpendicular magnetic anisotropy arising from the inverse of magnetostriction [Ref.24] because of the lattice mismatch between fcc-Cu (001) and fcc-Ni (001) having the lattice constants $a$ of 0.362 nm [Ref.25] and 0.352 nm [Ref.26], respectively. According to the studies on the magnetic properties for epitaxial Cu / Ni / Cu (001) sandwiches [Refs.27,28], the perpendicular magnetic anisotropy of this system comes from the *bulk* magnetoelastic anisotropy energy and *interface* magnetocrystalline anisotropy. The contribution of the former strain-induced effect is a major reason for the wide Ni thickness range exhibiting the spontaneous perpendicular magnetization from 3 nm to 12.5 nm [Ref. 28]. Since the strain plays an important role in the emergence of PMA not only for the Ni layer on the Cu substrate but also for the Ni / Cu superlattices [Refs.29,30], one may expect a larger PMA if a larger lattice strain can be induced in a Ni layer by using the nonmagnetic element having the larger lattice mismatch with Ni such as fcc-Pt ($a$ = 0.393) [Ref.31]. In 1990's, a few experimental studies reported on the fabrication of (111)-textured Ni / Pt superlattices with PMA [Refs.32-34]. However, no report had been made for Ni / Pt (001) epitaxial superlattices. Epitaxial or single crystal-like Ni / Pt superlattices are advantageous for the quantitative evaluation of their magnetic properties and the comparison to the theoretical calculations. In addition, taking into account the further investigation of transport and thermo-electric properties of Ni / Pt, Ni / Pt superlattices should be directly grown on a non-conductive substrate without any buffer layer materials, which is essential to examine the potential of Ni / Pt superlattices as spintronic and spin-caloritronic materials [Ref.35].

In this paper, we report the epitaxial growth of perpendicularly magnetized Ni / Pt (001) superlattices directly on a $SrTiO_3$ single crystal substrate, which was achieved by optimizing the film growth temperature. We show the structure and magnetic properties for the Ni / Pt (001) epitaxial superlattices, in particular the Ni layer thickness dependence of them. These experimental results are compared to the first-principles calculations, from which we were able to reveal that the major origin of the PMA is the strain effect for the Ni / Pt superlattice.

## 2. Experimental Procedure

The layers of $[Ni / Pt]_{\times N}$ were grown on a single crystal substrate employing an ultrahigh vacuum-compatible magnetron sputtering system with the base pressure below $2 \times 10^{-7}$ Pa. We chose the $SrTiO_3$ (100) single crystal substrate. We first deposited the Ni layer on the $SrTiO_3$ substrate followed by the Pt layer. The Ni layer thickness ($t$) was varied in the range from 1.0 nm to 6.0 nm while the Pt layer thickness was fixed at 1.0 nm. The total thicknesses of superlattices were designed to be approximately 20 nm by tuning the repetition number ($N$). The influence of substrate temperature ($T_s$) on the layer growth was investigated in order to achieve the epitaxial growth of [Ni / Pt]$_{\times N}$ on the SrTiO$_3$ (100) substrate, where $T_s$ was changed in the range from room temperature to 650ºC. Finally, using magnetron sputtering, a 2.0 nm-thick Al layer was deposited at room temperature on the [Ni / Pt]$_{\times N}$ layers as a capping layer.

The crystal orientation and the morphology of the layers were monitored by the *in-situ* observation using reflection high-energy electron diffraction (RHEED). Structural characterization was performed using the x-ray diffraction (XRD) with Cu-$K\alpha$ radiation and transmission electron microscope (TEM) together with the element analysis by the energy dispersive x-ray spectroscopy (EDX). Magnetic properties for the thin films were measured using a vibrating sample magnetometer (VSM) and a superconducting quantum interference device (SQUID) magnetometer (MPMS3-SQUID, Quantum Design, Inc.).

## 3. Experimental Results and Discussion

### 3.1 Growth temperature dependence

**Figure 1** displays RHEED images for [Ni / Pt] with $t = 3.0$ nm and $N = 5$, where $T_s$ was set at 200ºC, 400ºC and 650ºC. The diffraction patterns were observed just after the growth of the 5th Ni layers and the 5th Pt layers. At $T_s$ = 200ºC, the spots and the non-periodic streaks are observed, indicating that the Ni / Pt layers are not epitaxially grown on the SrTiO$_3$ (100) substrate. The diffraction patterns at $T_s$ = 200ºC also suggests that the preferential crystallographic orientation is the [111] direction for $T_s$ = 200ºC. As $T_s$ is increased to 400ºC, the diffraction patterns are drastically changed and the sharp streaks are observed. These streak patterns mean the epitaxial growth of Ni / Pt layers with the (001) plane. In addition, the 5th Ni layer and the 5th Pt layer exhibited different diffraction patterns due to the different surface morphologies, surface reconstructions and lattice constants for the Ni and Pt layers. These different diffraction patterns between the Ni and Pt layers imply that the intermixing between Ni and Pt layers are not remarkable at $T_s$ = 400ºC. However, the further increase in $T_s$ leads to the significant intermixing. Both the 5th Ni layer and the 5th Pt layer grown at $T_s$ = 650ºC show the similar diffraction patterns even though the streaks become sharper than those for $T_s$ = 400ºC. We consider that $T_s$ = 650ºC is too high to maintain the layered structure without intermixing. The intermixing at 650ºC is understood from the phase diagram of Ni-Pt binary alloy [Ref.36]. Since the liquidus is located around 1500ºC for the equiatomic composition of Ni-Pt, 650ºC is high enough for the induction of atomic diffusion.

The XRD profiles for [Ni / Pt] with $t$ = 3.0 nm and $N$ = 5 are shown in **Fig. 2** ((a) $T_s$ = 200ºC, (b) 400ºC, and (c) 650ºC). The reflections of Pt 111 and Ni 002 have the highest intensities for the samples at $T_s$ = 200ºC and 400ºC, respectively, in which the clear satellite peaks appear around the Pt 111 and Ni 200. The appearance of satellite peaks indicates the formation of layered structure. As mentioned above for the RHEED observation, the sample at $T_s$ = 200ºC is the non-epitaxial Ni / Pt with the (111) preferential orientation and the sample at $T_s$ = 400ºC is the epitaxial Ni / Pt with the (001) orientation. Thus, the appearances of Pt 111 and Ni 002 for $T_s$ = 200ºC and 400ºC, respectively, in the XRD profiles are consistent with the RHEED observation. In this study, however, Ni 111 and Pt 200 reflections were not identified because those peak angles, $2\theta$ = 44.6º for Ni 111 and $2\theta$ = 46.2º for Pt 200, were overlapped with the large peak from the SrTiO₃ substrate. In contrast to the samples at $T_s$ = 200ºC and 400ºC, the NiPt 002 appears and no satellite peak is observed at $T_s$ = 650ºC. We consider that alloying was promoted at $T_s$ = 650ºC, which is also consistent with the RHEED observation for $T_s$ = 650ºC.

**Figure 3** shows (a) the high-resolution TEM image and (b), (c) EDX mappings for Ni and Pt of [Ni / Pt] with $t$ = 3.0 nm and $N$ = 5 grown at $T_s$ = 400ºC. The initial Ni layer forms islands with the flat surface, and the subsequent Pt layer starts the layer growth on the Ni islands. As the layer number is increased, the layered structure becomes well-defined. From the structural characterization by RHEED, XRD and TEM, it is confirmed that the (001)-epitaxially grown Ni / Pt superlattice is achieved on the SrTiO₃ (001) substrate by optimizing the growth temperature such as $T_s$ = 400ºC.

**Figure 4** displays the magnetization curves for the [Ni / Pt] with $t$ = 3.0 nm and $N$ = 5 grown at (a) $T_s$ = 200ºC, (b) 400ºC, and (c) 650ºC. The red curves denote the magnetization curves measured with in-plane magnetic field (IP curve) while the blue curves denote those measured with out-of-plane magnetic field (OPP curve). The measurements were done at room temperature. In this study, the value of magnetization ($M$) was defined as the detected magnetic moment per the unit volume of Ni layers. In the case of $T_s$ = 200ºC, the easy magnetization axis lies in the film plane, which is attributable to the non-epitaxial Ni / Pt at $T_s$ = 200ºC. On the other hand, $T_s$ = 400ºC leads to the high squareness of out-of-plane magnetization curve, indicating that the Ni / Pt at $T_s$ = 400ºC possesses the PMA larger than the shape anisotropy. The further increase in $T_s$ up to 650ºC gives rise to the in-plane easy magnetization axis again. The disappearance of PMA results from the collapse of layered structure as explained in **Fig. 2(c)**.

In summary, $T_s$ = 400ºC is the adequate growth temperature for (001)-oriented epitaxial growth as well as the formation of layered structure, leading to the induction of PMA overcoming the shape anisotropy. The origin of PMA will be discussed in Sec. 3.3. Hereinafter, $T_s$ is fixed at 400ºC.

*3.2 Ni layer thickness dependence*

In this subsection, we show the $t$ dependence of structure and magnetic properties of [Ni / Pt] grown at $T_s = 400°C$, which allows us to reveal the origin of PMA for the (001)-epitaxially grown Ni / Pt superlattice. **Figure 5** shows (a) out-of-plane and (b) in-plane XRD profiles for $t = 1.5, 2.0, 3.0, 4.0, 5.0,$ and $6.0$ nm together with the profile for the 20.0 nm-thick Ni thin film grown on the SrTiO$_3$ (001) substrate. In the out-of-plane XRD profiles, the main reflections come from Ni 002 and clear satellite peaks are observed. As mentioned in the previous subsection, Pt 200 is not identified because of the overlap with the large substrate peaks. One sees that the $2\theta$ angles of main and satellite peaks monotonically shift with increasing $t$. As in the case of Ni 002, the Ni 200 reflection seen in the in-plane XRD profiles shows the gradual shift with $t$. **Figure 6** summarizes (a) the lattice constants of $a$- and $c$-planes, (b) the value of $c / a$, and (c) the superlattice period ($D$) as a function of $t$. $D$ was calculated by the following equation: $(2/\lambda) \sin \theta_n = 1/d \pm n/D$ for the $n$th satellite peaks with the x-ray wavelength ($\lambda$) and the lattice spacing ($d$). As shown in **Fig. 6(a)**, the value of $a$ is larger than that for bulk Ni [Ref.26] whereas the value of $c$ is smaller than that for bulk Ni. These tendencies become remarkable as $t$ is reduced. As a result, the value of $c / a$ is decreased down to 0.90 at $t = 1.2$ nm. This means that a larger tensile strain exists in the film plane for smaller $t$, and the lattice strain is relaxed as $t$ is increased, leading to the values of $a$ and $c$ approaching to the bulk lattice constant.

**Figure 7** shows the magnetization curves for the [Ni / Pt] with (a) $t$ = 1.2, (b) 1.3, (c) 1.5, (d) 2.0, (e) 4.0, and (f) 5.0 nm, which were measured at room temperature. All the films except for $t$ = 1.2 nm possess the PMA, resulting in the easy magnetization axis normal to the film plane. The effective uniaxial magnetic anisotropy constant ($K_{eff}$) corresponds to the area enclosed between the OPP and IP curves. As $t$ is increased from 1.3 to 4.0 nm, the saturation field of IP curve gradually increases, indicating the increase in $K_{eff}$. The further increase in $t$ up to 5.0 nm gives rise to the reduction of $K_{eff}$. The values of $K_{eff}$, saturation magnetization ($M_s$) and uniaxial magnetic anisotropy constant ($K_u = K_{eff} + 2\pi M_s^2$) as a function of $t$ are plotted in **Figs. 8(a), 8(b) and 8(c)**. The value of $K_{eff}$ is significantly increased in the range of 2.0 nm ≤ $t$ ≤ 4.0 nm. This means that there exists an adequate $t$ region for enhancing the PMA. $M_s$ is monotonically decreased as $t$ is decreased from 6.0 to 2.0 nm. Below $t$ = 2.0 nm, $M_s$ is steeply decreased and the sample for $t$ = 1.0 nm does not exhibit the spontaneous magnetization at room temperature. The $t$ dependence of $K_u$ is similar to that of $K_{eff}$. One sees that the [Ni / Pt] samples with $t$ larger than 4.0 nm also possess the moderate $K_u$. The maximum $K_u$ is $1.9 \times 10^6$ erg/cm$^3$ for $t$ = 4.0 nm.

The remarkable points observed in the $t$ dependence of $K_u$ are the broad thickness region for high $K_u$ and the drastic reduction of $K_u$ below $t$ = 2.0 nm. These features are different from the conventional metallic superlattices showing perpendicular magnetization thanks to the interface magnetic anisotropy, such as Co / Pd [Refs.10-12], Co / Pt [Refs.11-13], and Co / Ni [Refs.14-19]. In contrast to these systems, the broad thickness region for high $K_u$ was reported also for the Cu / Ni / Cu (001) sandwich structures [Ref.28]. According to the previous work on the Ni / Cu system [Ref.28], the strain dependent magnetic surface anisotropy plays a major role for inducing the PMA in the Ni / Cu superlattices. In that case, $K_{eff}$ is phenomenologically described by

$$K_{eff} = -2\pi M_s^2 + 2\left(B_1 + \frac{B_s}{t}\right)e_0(t) + \left(K_1 + \frac{2K_s}{t}\right), \quad (1)$$

where $B_1$ is the first-order cubic bulk magnetoelastic coupling coefficient, $B_S$ is the surface magnetoelastic coupling coefficient, $K_1$ is the first-order cubic magnetocrystalline anisotropy energy, and $K_s$ is the surface magnetic anisotropy energy. $e_0(t)$ represents the average in-plane biaxial misfit strain, which is given by $e_0(t) = \eta(t_c/t)$ using the form of the average strain [Ref.37], where $\eta$ is the film-substrate lattice mismatch and $t_c$ is the thermodynamic critical thickness. With the assumption that $K_1$ is negligibly small, Eq. (1) can be transformed into

$$K_{eff}t = -2\pi M_s^2 t + 2(B_1 t_c \eta + K_s) + \left(\frac{2B_s t_c \eta}{t}\right). \quad (2)$$

In **Fig. 8(d)**, $K_{eff} t$ as a function of $t$ is plotted. The experimental data at $t \geq 2.0$ nm, where the value of $M_s$ keeps the almost constant value of 400 emu/cm$^3$, was fitted with Eq. (2) using $B_1 = 6.2 \times 10^7$ erg/cm$^3$ [Ref.28]. In this study, we assumed the value of $\eta$ as $\eta = (a_f - a_s) / a_s$ with the lattice constants of film ($a_f$) and substrate ($a_s$). If $\eta$ is calculated using the bulk lattice constants of Ni ($a = 0.352$ nm) and Pt ($a$

= 0.393 nm), $\eta$ is obtained to be 0.1. In this case, the Pt layers are regarded as a very solid layer like a substrate. However, this idea may not be appropriate because the lattice constant of Pt is also influenced by the formation of interface with the Ni layers. This means that the actual value of $\eta$ must be much lower than 0.1. Unfortunately, it is difficult to strictly determine the value of $\eta$. Then, based on the previous study on Ni / Cu [Ref.28], in which $\eta$ was set at 0.026, we fitted the present experimental result with $\eta$ = 0.026 and $t_c$ = 1.8 nm. As a result, $K_s$ = 0.15 ± 0.04 erg/cm$^2$ and $B_s$ = -10.4 ± 3.0 erg/cm$^2$ were obtained. Those are of the same order as the values reported for the Ni / Cu system [Ref.28]. Because of the uncertainty of $\eta$ as discussed above, it is hard to quantitatively discuss the values of $K_s$ and $B_s$. That uncertainty also may be a reason why the steep change in $K_{eff} t$ at 4.0 nm ≤ $t$ ≤ 5.0 nm is not reproduced by the calculation. We however emphasize that Eq. (2) qualitatively explains the experimental tendency, which strongly suggests that the strain effect, *i.e.* the value of $B_1$, largely contributes to the emergence of PMA rather than $K_s$. In the next subsection, the effect of the lattice strain on the magnetic anisotropy will be discussed based on the comparison between the experimental results and theoretical calculation.

Although the increased $K_u$ in the thickness range of 2.0 nm ≤ $t$ ≤ 4.0 nm is attributable to the strain effect, the lattice strain cannot explain the remarkable reduction of $K_u$ below $t$ = 2.0 nm because the Ni lattice is significantly distorted even at $t$ < 2.0 nm. In **Fig. 8**, one may be aware that the reduction

Page 13

of $K_u$ is accompanied by the reduction of $M_s$. In order to understand the reason for the remarkable reduction of $M_s$ at $t < 2.0$ nm, we measured the measurement temperature ($T$) dependence of $M$ for $t =$ 1.0, 1.5, 2.0, 3.0 and 4.0 nm as shown in **Fig. 9**. These $M$-$T$ curves suggest the Curie temperature is gradually decreased with decreasing $t$ and becomes lower than room temperature for $t = 1.0$ nm. Consequently, we find that the remarkable reduction of $M_s$ at $t < 2.0$ nm originates from the decrease in the Curie temperature for the thin Ni layers.

*3.3 Theoretical calculations*

In this subsection, the effect of the lattice strain on the magnetic anisotropy is discussed based on the first-principles calculation results. The first-principles calculations were performed by using Vienna ab initio Simulation Package [Ref.38] with the Generalized Gradient Approximation parameterized Perdew, Burke, and Ernzerhof [Ref.39] and projector augmented wave potentials [Ref.40]. The wave functions were expanded in a plane wave basis set up to a cutoff kinetic energy of 500 eV. The cell volume and all of atomic positions were relaxed within the constraint of the fixed in-plane lattice constant. The magnetic anisotropy energy (MAE) was obtained using the magnetic force theorem method. The 24 x 24 x 24 and 24 x 24 x 1 k-point mesh were used for the evaluation of the MAE for the Ni bulk and the multilayer consisting of Ni 17 monolayers (MLs) and Pt 17MLs. First, we examined the MAE for the Ni bulk induced by the tetragonal lattice distortion as shown in **Fig. 10(a)**. In the present calculation, the positive MAE means that the easy magnetization axis lies along the *c*-axis of distorted Ni. The positive MAE is induced for the Ni bulk at $c/a < 1$, i.e. by the in-plane tensile strain. The MAE is proportional to the orbital moment anisotropy for the 3*d* transition metals as derived by Bruno [Ref.41]. The orbital magnetic moment is increased as the tensile strain is induced while the spin magnetic moment is decreased (see **Figs. 10(b) and 10(c)**). One sees that the orbital magnetic moment magnetized along the [001] direction is enhanced by the tensile strain as shown in **Fig. 10(c)**, resulting in the induction of PMA. In order to obtain the insight of PMA, we estimated the MAE contribution from the 2nd-order perturbation of the spin-orbit coupling (see **Fig. 10(d)**) [Refs.41,42]. The PMA mainly comes from the spin-conserving term between the minority spin states whereas the spin-flipping contribution is negligible for the MAE of Ni bulk.

Next, we calculated the MAE for the Ni/Pt interface. The layer resolved MAE for the Ni/Pt interface is shown in **Fig. 11**, where the in-plane lattice constant of Ni matches that of Pt. **Figure 11(a)** illustrates the model of calculation, and **Figs. 11(b) and 11(c)** are the position dependence of MAE calculated with the in-plane lattice constant of $a = 0.352$ nm and $a = 0.372$ nm, respectively. The interfacial Ni and Pt layers show the PMA regardless of the in-plane lattice parameter. However, the 2nd and 3rd Ni MLs away from the interface exhibit the in-plane magnetic anisotropy in the case of in-plane lattice constant for the Ni bulk (**Fig. 11(b)**). **Figure 12** shows the local density of states for the Ni atom in Ni/Pt multilayer. At the interfacial Ni atom, $d_{x^2-y^2}$ and $d_{xy}$ states exist below and above the Fermi level, respectively, and the spin-orbit coupling matrix element of $\langle d_{x^2-y^2}|\ell_z|d_{xy}\rangle$ contributes to the PMA. For the 2nd and 3rd layers, on the other hand, $d_{3z^2-r^2}$ state is increased near the Fermi level compared to $d_{x^2-y^2}$ state, and the matrix element of $\langle d_{3z^2-r^2}|\ell_x|d_{xz}\rangle$ contributes to in-plane magnetic anisotropy. For the multilayer with the in-plane lattice constant of Pt bulk (**Fig. 11(c)**), all the Ni layers from the 4th to 14th ML show the large PMA induced by the tensile tetragonal distortion. The PMA of the Ni / Pt superlattice is attributed to both the non-negligible interfacial contribution and the major bulk contribution induced by lattice distortion. This is different from the other Ni-based superlattices such as Ni / Au and Ni / Pd, in which the interfacial contribution is negative, i.e. $K_s$ < 0, or negligibly small [Ref.20]. As mentioned in the analysis of **Fig. 8(d)**, we have found that the strain effect largely contributes to the emergence of PMA for the Ni / Pt superlattices. At the same time, we experimentally evaluated the non-negligible interface magnetic anisotropy energy. Therefore, the above first-principles calculation results are qualitatively consistent with the experimental results.

**4. Summary**

We investigated the optimum film growth conditions to achieve the epitaxial growth of the perpendicularly magnetized Ni / Pt (001) superlattices directly on a SrTiO$_3$ (001) single crystal substrate. We found that $T_\text{s}$ = 400ºC was the adequate growth temperature for (001)-oriented epitaxial growth as well as the formation of layered structure. This (001)-oriented epitaxial growth induced the PMA overcoming the shape anisotropy, resulting in the perpendicularly magnetized Ni / Pt. We obtained the high $K_\text{u}$ = 1.9 × 10$^6$ erg/cm$^3$ for $t$ = 4.0 nm. The Ni layer thickness dependence of structural parameters and magnetic properties clearly indicated that the strain effect largely contributes to the emergence of PMA. This experimental finding was supported by the first-principles calculation. The first-principles calculation also suggested the non-negligible contribution of interface magnetic anisotropy to the PMA, which was qualitatively consistent with the experimental results. The findings in this study will provide with the useful knowledge for developing a perpendicularly magnetized superlattice.

**Acknowledgement**

The authors thank T. Kubota and Y. Sakuraba for their valuable comments. The structural characterization was partly carried out at the Cooperative Research and Development Center for Advanced Materials, IMR, Tohoku University. This work was supported by the Grant-in-Aid for Scientific Research (S) (JP18H05246) and the GIMRT Program of the Institute for Materials Research, Tohoku University (Proposal No. 19K0506).

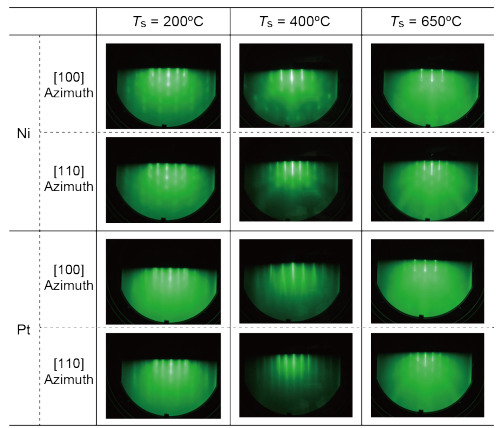

**Figure 1** Reflection high-energy electron diffraction images for [Ni / Pt] with $t$ = 3.0 nm and $N$ = 5, where $T_s$ was set at 200ºC, 400ºC and 650ºC. The diffraction patterns were observed just after the growth of the 5th Ni layers and the 5th Pt layers.

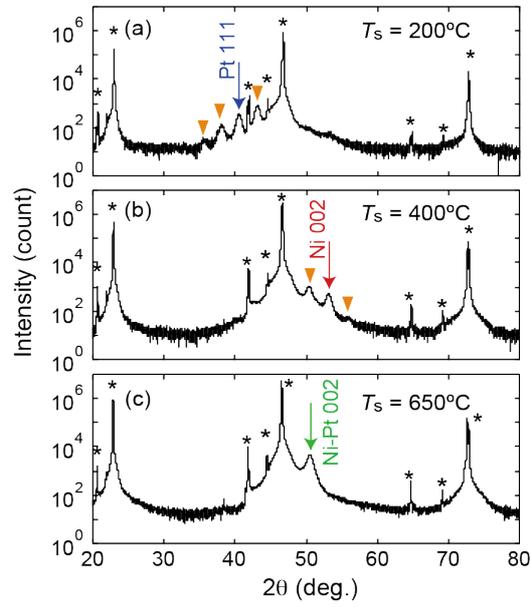

**Figure 2** X-ray diffraction profiles for [Ni / Pt] with $t$ = 3.0 nm and $N$ = 5 grown at (a) $T_s$ = 200ºC, (b) 400ºC, and (c) 650ºC. The asterisks denote the reflections from the SrTiO3 (001) substrates. The inverted triangles represent the satellite reflections.

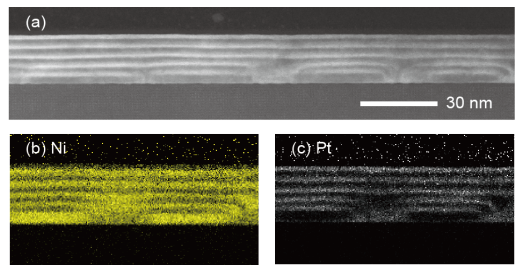

**Figure 3** (a) High-resolution transmission electron microscope image and (b) energy dispersive x-ray spectroscopy mappings for Ni and (c) Pt of [Ni / Pt] with $t$ = 3.0 nm and $N$ = 5 grown at $T_s$ = 400ºC.

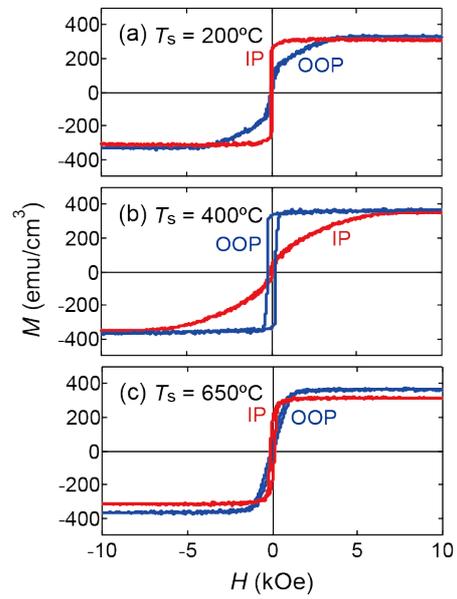

**Figure 4** Magnetization curves for the [Ni / Pt] with $t$ = 3.0 nm and $N$ = 5 grown at (a) $T_s$ = 200ºC, (b) 400ºC, and (c) 650ºC. The red curves denote the magnetization curves measured with the in-plane magnetic field (IP) while the blue curves denote those measured with the out-of-plane magnetic field (OPP). The measurement was done at room temperature.

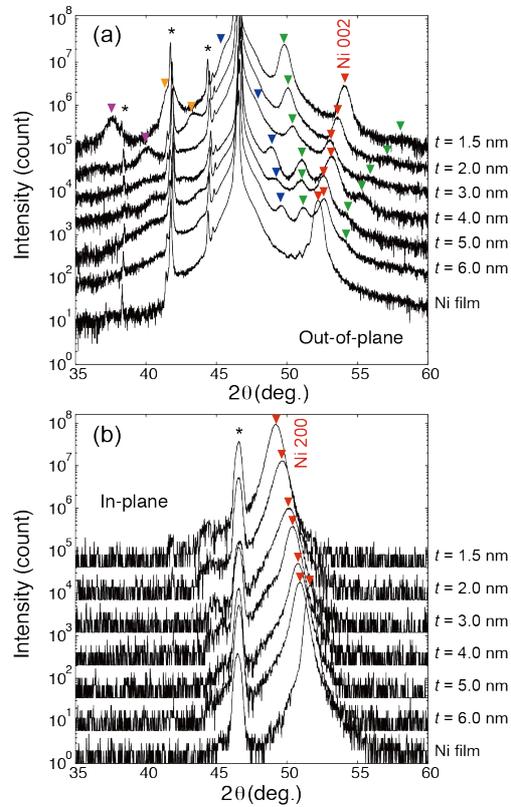

**Figure 5** (a) out-of-plane and (b) in-plane x-ray diffraction profiles for the [Ni / Pt] with $t$ = 1.5, 2.0, 3.0, 4.0, 5.0, and 6.0 nm, which were grown at $T_s$ = 400ºC, together with the profile for the 20-nm thick Ni thin film grown on the SrTiO$_3$ (001) substrate. The asterisks denote the reflections from the SrTiO3 (001) substrates. In (a), the inverted triangles except for the red triangles represent the satellite reflections of Ni 002 peaks.

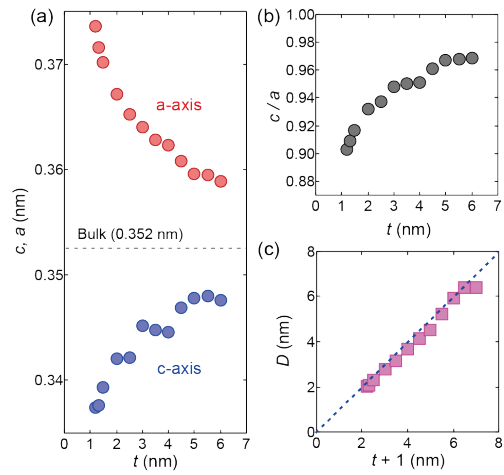

**Figure 6** (a) Lattice constants of *a*- and *c*-planes, (b) the value of $c/a$, and (c) the superlattice period (*D*) as a function of *t* for the [Ni / Pt] grown at $T_s = 400°C$.

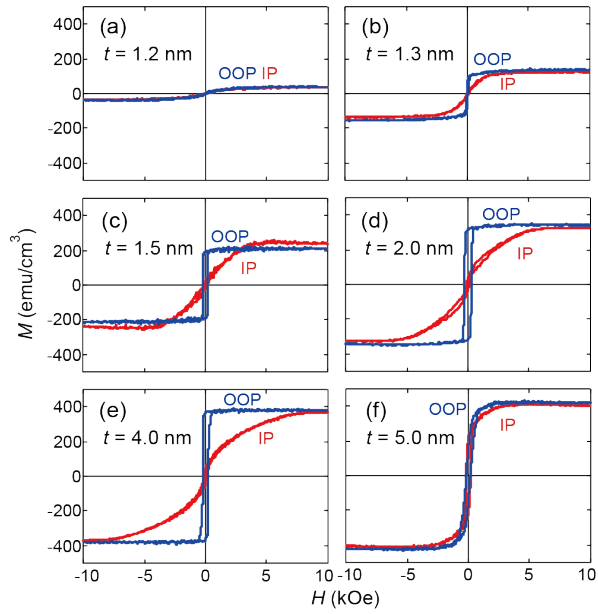

**Figure 7** Magnetization curves for the [Ni / Pt] with (a) $t$ = 1.2, (b) 1.3, (c) 1.5, (d) 2.0, (e) 4.0, and (f) 5.0 nm, which were grown at $T_s$ = 400ºC. The red curves denote the magnetization curves measured with the in-plane magnetic field (IP) while the blue curves denote those measured with the out-of-plane magnetic field (OPP). The measurement was done at room temperature.

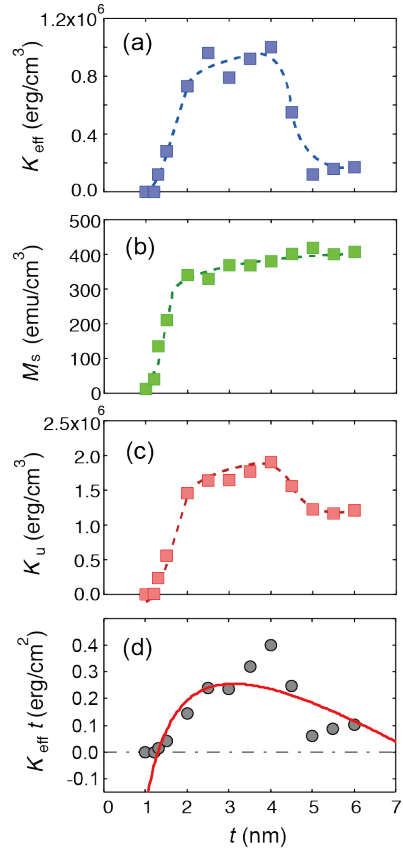

**Figure 8** (a) Effective uniaxial magnetic anisotropy constant ($K_{eff}$), (b) saturation magnetization ($M_s$), (c) uniaxial magnetic anisotropy constant ($K_u$), (d) $K_{eff} t$ and as a function of $t$ for the [Ni / Pt] grown at $T_s$ = 400ºC. The solid curve shown in (d) is the result of fitting using Eq. (2).

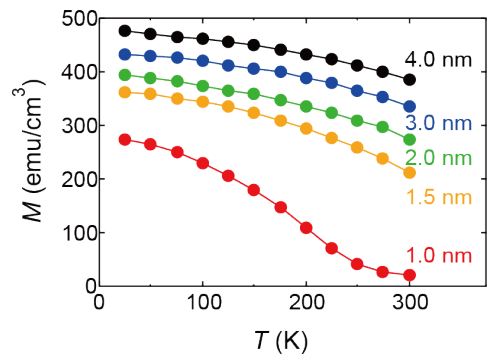

**Figure 9** Measurement temperature (*T*) dependence of magnetization (*M*) for the [Ni / Pt] with *t* = 1.0,

1.5, 2.0, 3.0 and 4.0 nm, which were grown at $T_s$ = 400ºC.

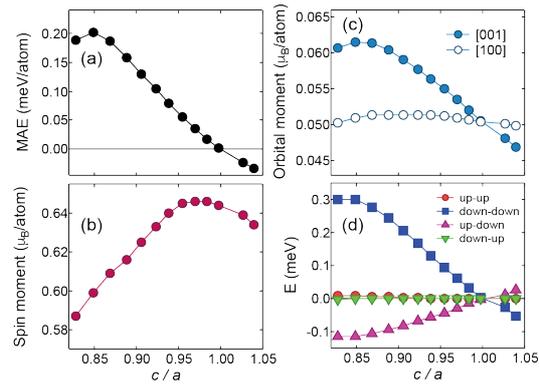

**Figure 10** (a) Magnetic anisotropy energy (MAE), (b) spin moment, (c) orbital moments for [001] and [100] directions and (d) energies ($E$) for the spin-conserving process (up-up or down-down) and the spin flip process (up-down or down-up) as a function of $c/a$ for the bulk Ni.

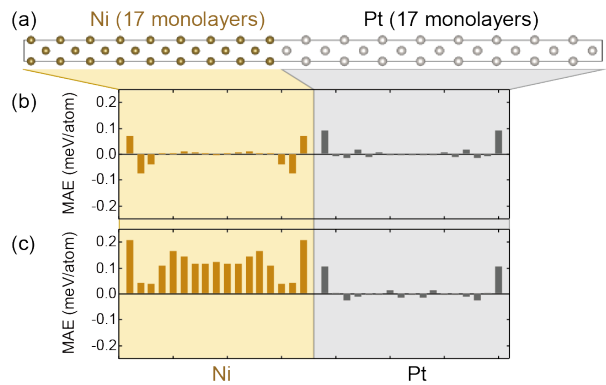

**Figure 11** (a) Schematic illustration of model for the first-principles calculation, which consists of Ni 17 monolayers and Pt 17 monolayers. (b) Position dependence of MAE calculated with the in-plane lattice constant of $a = 0.352$ nm and (c) $a = 0.372$ nm.

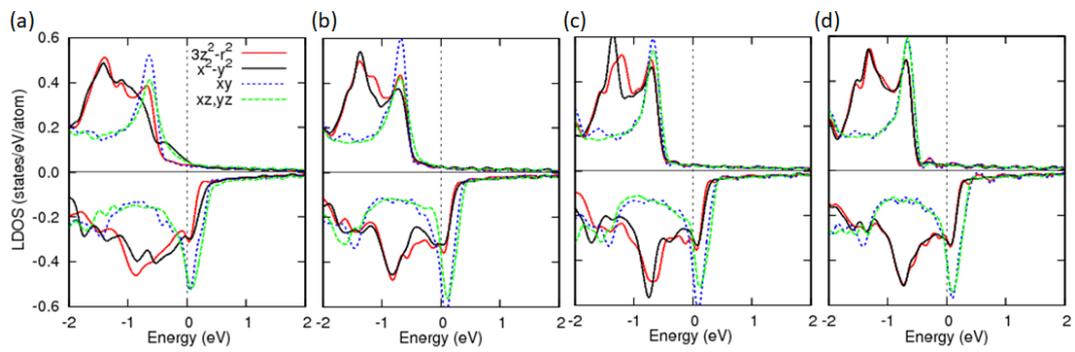

**Figure 12** (a) Local density of states for 1st, (b) 2nd, (c) 3rd, and (d) 9th Ni atomic layer from the Ni / Pt interface, where the in-plane lattice constant was set at $a$ = 0.352 nm.